\documentstyle[12pt,amssym,aaspp4,flushrt]{article}

\lefthead{GEROYANNIS, PAPASOTIRIOU, & VALIAKA}
\righthead{VISCOUS EFFECTS IN WHITE DWARFS AND NEUTRON STARS}

\begin{document}

\title{VISCOUS DISSIPATIVE EFFECTS IN WHITE DWARFS\\
                  AND NEUTRON STARS}

\author{V. S. Geroyannis, P. J. Papasotiriou, and I. Valiaka}
\affil{Astronomy Laboratory, Department of Physics,\\
       University of Patras,GR--26500 PATRAS, GREECE;\\
       first author's e-mail: vgeroyan@physics.upatras.gr}

\begin{abstract}
When a differentially rotating magnetic star undergoes ``turn-over'', i.e., its magnetic symmetry axis is inclining at a gradually increasing angle with respect to its invariant angular momentum axis up to the perpendicular position (situation known as ``perpendicular rotator''), the poloidal magnetic field causes interchange of angular momentum via hydromagnetic Alfven waves propagating along the poloidal magnetic field lines and, accordingly, the resulting perpendicular rotator obtains uniform rotation. This dissipative mechanism is represented by the so-called ``turn-over viscosity''. For several white dwarf and neutron star models, we compute the energy dissipation due to the turn-over viscosity; we also compare this effect with the corresponding dissipative effects induced by the electron, ion, and turbulent viscosity.  
\end{abstract}


\section{Introduction}
According to the ``turn-over scenario'' (for details see \cite[hereafter G01, \S \, 3]{gero01}), the rotating magnetic star undergoes an axisymmetric ``early evolutionary phase'', during which the combined effect of rotation and (toroidal plus poloidal) magnetic field gives configurations with moment of inertia along the principal axis $I_3$ (coinciding with the rotation axis, the magnetic symmetry axis, and the invariant angular momentum axis, taken to be the $z$ axis), $I_{33}$, greater than the moments of inertia along the principal axes $I_1$ and $I_2$, $I_{11} = I_{22}$; i.e., $I_{33} > I_{11}$. At some particular time, however, which signals the beginning of a ``late evolutionary phase'', secular angular momentum loss due to the surface poloidal magnetic field, $B_s$, induces configurations with $I_{33}(L_x) = I_{11}(L_x)$ (relation valid for a particular angular momentum $L_x$) and, after that particular time, configurations with $I_{33}(L) < I_{11}(L)$ when $L < L_x$. Thus, a ``dynamical asymmetry'' is established in the sense that the moment of inertia along the rotation axis becomes less than the moments of inertia along the other two principal axes. This inequality is due to the action of the toroidal magnetic field, which tends to derive prolate configurations; while rotation and poloidal magnetic field tend to derive oblate configurations. 

Dynamically asymmetric configurations tend to turn over spontaneously and thus to rotate along axis with moment of inertia greater than $I_{33}$. So, they become ``oblique rotators''; remark that ``perpendicular rotator'' is an oblique rotator with ``turn-over angle'' (or ``obliquity angle'': the angle of the magnetic symmetry axis with respect to the invariant angular momentum axis), $\chi$, equal to $90 \arcdeg$ (the internal dynamics of an oblique rotator has been studied in detail by \cite[hereafter MT72]{mtak72}). 

The ``terminal'' nonaxisymmetric rigidly rotating model becomes perpendicular rotator on a ``turn-over timescale'', $t_{TOV}$, comparable with the ``period timescale'', $t_{P} = P_{now}/\dot{P}_{now}$; i.e., $t_{TOV} \sim t_{P}$. Although any intermediate oblique model with $\chi < 90 \arcdeg$ undergoes a complicated rotation along axis passing through the center of mass but not coinciding with one of its principal axes (MT72, \S\S \, 2--3), the terminal model rotates along its $I_1$ principal axis, which axis is now coinciding with the invariant angular momentum axis taken, that is, the $z$ axis. Apparently, its angular momentum vector remains equal to that of the starting model (MT72, \S \S \, 1--2), while its angular velocity vector obtains same direction with that of the starting model (coinciding with the angular momentum direction) but different magnitude. 

During the turn-over phase, the turn-over angle increases spontaneously up to $90 \arcdeg$ on timescale $t_{TOV}$, since the rotational kinetic energy of the model decreases from a higher level with turn-over angle $\chi \gtrsim 0 \arcdeg$ to a lower level with turn-over angle $\chi \simeq 90 \arcdeg$. At this level, the model reaches the state of least energy consistent with its prescribed angular momentum and magnetic field. Accordingly, the excess energy due to the differential rotation field of motions, defined by the angular velocity component $\Omega_3$ along the (spontaneously turning over) $I_3$ principal axis (coinciding, in turn, with the magnetic symmetry axis), is dissipated down to zero under the action of turbulent viscosity in the convective regions of the model (MT72, \S \S \, 3--4). Thus, differential rotation along the $I_3$ principal axis decreases down to zero and all of its effects disappear.

On the other hand, it is difficult for an oblique model to sustain differential rotation along its $I_1$ principal axis mainly due to the destructive action of the poloidal magnetic field (see, e.g., \cite[hereafter MMT90, \S\S \, 1--3]{mmta90}; also, \cite[hereafter M92, \S\S \, 1, 5-6]{moss92}). Indeed, when a model has magnetic field symmetry axis inclined with respect to the rotation axis, nonuniform rotation in the electrically conducting stellar material shears the field and generates magnetic torques. Thus, there is a competition between the efforts of the magnetic stresses to remove rotational nonuniformities, and those of the rotational velocities to bury and destroy magnetic flux. If the magnetic field and the electrical conductivity have appropriate values (see MMT90, \S \, 4; also M92, \S \, 6), then the magnetic field prevails and removes the nonuniformities of rotation. Hence, the terminal model rotates rigidly along its $I_1$ principal axis with angular velocity $\Omega_1$. 

\section{The turn-over viscosity}
The turn-over timescale, $t_{TOV}$, is given by (G01, eq. [41])
\begin{equation} \label{eq:ttov}
t_{TOV} \simeq 
  \frac{D\!E_{DRD}}{V}
  \left \langle \Omega_3 \right \rangle^{-2}
  \left \langle \frac{H^2}{8 \pi \varrho a^2} \right \rangle^{-2} 
  \left \langle \frac{1}{3} \, \frac{\varrho V_t}{\Lambda_t} \right \rangle_{sz}^{-1}
  \left \langle \frac{\Omega_3^2 r^2}{a^2} \right \rangle_{sz}^{-2}
  \left \langle \Lambda_t \right \rangle_{sz}^{-2} 
  {\rm \, ,} 
\end{equation}
where the subscript $sz$ denotes that averages calculated over the surface zone with base at the transition layer, $\xi_s$, and top at the boundary, $\xi_1$, of the star (G01, eqs. [13]--[18]); averages without subscripts denote global averages. This $t_{TOV}$ estimate results as the mean energy per unit volume available for dissipation, $D\!E_{DRD} / V$ ($D\!E_{DRD}$ is the excess rotational kinetic energy due to differential rotation and $V$ is the volume of the configuration), divided by the mean energy dissipation  during the turn-over due to turbulent viscosity in the convective surface zone per unit volume and per unit time (i.e., the product of the last 5 terms in the right-hand side of eq. [\ref{eq:ttov}]; for details see MT72, eqs. [7], [49]--[50]). The most quantities in equation (\ref{eq:ttov}) are calculated under the assumption of slightly declining from sphericity. In particular, the density is approximated by $\varrho(\xi,\nu) \simeq \varrho(\xi)$, the pressure is approximated by $P(\xi,\nu) \simeq P(\xi)$, the local sound speed, $a$, is approximated by 
\begin{equation}  
a(\xi,\nu) \simeq a(\xi) = \sqrt{\frac{P(\xi)}{\varrho(\xi)}} {\rm \, ,}
\end{equation}
the turbulent mixing length, $\Lambda_t$, results from the ``mixing-length theory'' (MLT; see, e.g., \cite[hereafter CM91]{cmaz91}) as $\Lambda_t = A_P \, h_P$ (CM91, \S \S \, 2.8, 3.2), where the fine-tuning parameter $A_P$ is taken equal to unity and $h_P$ is the pressure scaleheight, 
\begin{equation}
\Lambda_t(\xi,\nu) \simeq \Lambda_t(\xi) \simeq 
          h_P(\xi) = A_{MLT} \, \, \frac{P(\xi)}{\varrho(\xi) g(\xi)} = 
          A_{MLT} \, \, \frac{P(\xi)(\alpha \xi)^2}{G \varrho(\xi) M(\xi)} 
                                                                {\rm \, ;} 
\end{equation}
$A_{MLT}$ is the standard MLT fine-tuning parameter taken to be of order $10^{-1}$, $M(\xi)$ is the mass inside a sphere of radius $\xi$, and the turbulent velocity, $V_t$, is approximated by (see, e.g., \cite[hereafter B90]{bric90}, eq. [5]; in combination with CM91, eq. [38])
\begin{equation}
V_t(\xi,\nu) \simeq V_t(\xi) \simeq 
             \sqrt{\frac{3 P(\xi)}{\varrho(\xi)}} {\rm \, .}
\end{equation}

An interesting question is what kind of globally acting ``turn-over viscosity'', $\mu_{TOV}$, can substitute the locally acting ``turbulent viscosity'', $\mu_t$, so as to induce energy dissipated over the time $t_{TOV}$, $D\!E_{TOV}$, equal to the difference $D\!E_{DRD} = T_{xx} - T_{RR}$ of the rotational kinetic energies of the starting and the terminal models. The reason for seeking a global viscosity, instead of the local turbulent viscosity, is that differential rotation, which causes this kind of viscosity, is a global property of the model.
We can first substitute the local turbulent velocity field, $V_t$, by a global ``turn-over velocity field'', $V_{TOV} \sim R / t_{TOV} \sim V_{mp}$ (the latter symbol denotes the mean speed of the magnetic poles on trajectories coinciding with circumference quadrants). Then the turbulent viscosity (see, e.g., B90, eq. [8])
\begin{equation}        
\mu_t(\xi) = \frac{1}{3} \varrho(\xi) V_t(\xi) A_P h_P(\xi) \simeq
             \frac{1}{3} \varrho(\xi) 
             \left\langle V_t \right\rangle_{sz}
             \left\langle h_P \right\rangle_{sz}
\end{equation}
can be substituted by the turn-over viscosity
\begin{equation}        
\mu_{TOV}(\xi) \simeq
   \frac{1}{3} \varrho(\xi) 
   \left\langle V_{TOV} \right\rangle A_{TOV}^h 
   \left\langle h_P \right\rangle \simeq
   \frac{1}{3} \varrho(\xi) \frac{A_{TOV}^h R^2}{10^2 \, t_{TOV}} {\rm ,}   
\end{equation}
(since for our models $\left \langle h_P \right \rangle \sim R / 10^2$) where $A_{TOV}^h$ is a fine-tuning parameter. Furthermore, the energy dissipated due to such turn-over viscous friction per unit time, $D_{TOV}$, is given by (see, e.g., \cite{gsid95}, eq. [7])
\begin{equation} \label{eq:DTOV}
D_{TOV} = 4 \pi \alpha^3 \Omega_{3c}^2 
   \int_0^1{
            \int_0^{
                    \xi_1}{\mu_{TOV}(\xi) \, \xi^4 (1 - \nu^2) 
            \left( \frac{d\omega}{ds} \right)_{s=s(\xi,\nu)}^2 d\xi
                   }
           } d\nu {\rm ,}
\end{equation}  
and assuming that $D_{TOV}$ remains constant over the time $t_{TOV}$ the energy, $D\!E_{TOV}$, dissipated due to such viscous friction over the time $t_{TOV}$ is equal to $t_{TOV} D_{TOV}$; so, this equality leads to the determination of the fine-tuning parameter $A_{TOV}^h$. 

A second step then is to develop a phenomenological model involving further details on the time evolution of the turn-over process (G01, \S \, 6). In the so-called ``mixer-mixture model'' (MMM), we assume that the turning over configuration acts as a ``mixer'' on the ``mixture'' configuration, which remains almost axisymmetric with respect to the invariant angular momentum axis, i.e., the $z$ axis (the meaning of ``almost axisymmetric configuration'' is that the angle $\gamma$ between its angular velocity and its invariant angular momentum remains small: $\gamma < 1 \arcdeg$, say; apparently, $\gamma$ becomes zero after the termination of the turn-over; in the following, we shall assume that such a deviation of the mixture configuration from axisymmetry is negligible and we shall treat it as being axisymmetric with respect to the $z$ axis). The mixer consumes its excess rotational kinetic energy due to differential rotation in the process of homogenizing the rotation of the mixture. In particular, when viewed in the mixture's system of reference, the mixer generates a velocity field, which carries mass elements across cylinders of different angular velocities and momenta. This is due to the fact that the rotational circular orbit of a mass element of the mixer, as seen in the mixer's system of reference, crosses several cylinders with different angular velocities, as seen in the mixture's system of reference. Thus, there is an interchange of angular momentum, which leads to uniform rotation of the mixture. 

For (1) the turn-over angle $\chi$ (with $\nu_{\chi} = \cos(\chi)$), (2) the central angular velocity $\Omega_{3c}$ of the mixer, and (3) the differential rotation reduction factor $F_r$ of the mixer, we adopt the time dependent scheme (G01, eqs. [52]--[54]) 
\begin{equation} \label{eq:tdep}
Q(Q_{ini},t) = \left(
             \frac{1+e^{(\kappa -1)\delta}}
                  {1+e^{(\kappa - \nu_L(t))\delta}} 
       \right) Q_0(Q_{ini}) -
       \left(
             \frac{1+e^{(\kappa -1)\delta}}
                  {1+e^{\kappa \delta}} 
       \right) Q_0(Q_{ini}) {\rm ,}
\end{equation}     
where the so-called ``threshold parameter'', $\kappa$, is taken equal to $\kappa = \cos(45 \arcdeg) \simeq 0.707$, the so-called ``delta parameter'', $\delta$, is chosen equal to $\delta \simeq 5$, and the auxiliary function $\nu_L(t)$ varies linearly with the time in the interval $[0,t_{TOV}]$,
\begin{equation} \label{eq:nuLt}
\nu_L(t) = \left \{ 
           \begin{array}{ll}
           1 - \frac{t}{t_{TOV}} & {\rm if} \ t \leq t_{TOV} \\
           0                     & {\rm if} \ t > t_{TOV}
           \end{array}
           \right. {\rm ;}
\end{equation}
When $\delta$ is relatively large ($\delta = 50$, say), the second term in the right-hand side of equation (\ref{eq:tdep}) becomes negligible with respect to its counterpart, and $Q_0(Q_{ini})$ can be simply set equal to the initial value(s) $Q_{ini}$, $\nu_{\chi[ini]} \lesssim 1$, $\Omega_{3c[ini]} = \Omega(L_{xx})$, and $F_{r[ini]} = 1.00, \; {\rm or} \; 0.50$. However, as our numerical results show, the present study requires $\delta \simeq 5$; since then the second term in the right-hand side of equation (\ref{eq:tdep}) cannot be neglected, $Q_0(Q_{ini})$ must be normalized to the value   
\begin{equation}
Q_0(Q_{ini}) = \frac{Q_{ini}}{1 - \left(
                         \frac{1+e^{(\kappa -1)\delta}}
                              {1+e^{\kappa \delta}}
                         \right)} {\rm \, ,}
\end{equation}
so that $Q(Q_{ini},0) = Q_{ini}$.
 
In the framework of MMM, we use a simplifying description according to which the turn-over speed at a particular point is proportional to the relative difference of the linear speeds due to differential rotation, $v_{mxr}(\xi, \nu_{mxr},t)$ and $v_{mxt}(\xi,\nu_{mxt},t)$, in the mixer and in the mixture --- where the rotation of the latter results as superposition of a gradually established rigid rotation about the $z$ axis with angular velocity $\Omega_{z}$ (which is not involved in the calculation of the turn-over speed) and of a differential rotation about the $z$ axis identical to the differential rotation of the mixer about the $I_3$ principal axis; for simplicity, we shall use the same symbols for the central angular velocity $\Omega_{3c}$ and the angular velocity field $\Omega_3$ for the differential rotations of both the mixer and the mixture, keeping in mind that the only difference concerns their rotation axes --- (G01, eq. [55]), 
\clearpage     
\begin{eqnarray} \label{eq:vtov}
V_{TOV} & = & V_{mp} 
              \left( 
              \frac{\left| v_{mxt}(\xi,\nu_{mxt},t) - 
                           v_{mxr}(\xi,\nu_{mxr},t) \right|}
                   {v_{mxt}(\xi,\nu_{mxt},t)}
              \right)  = \nonumber \\
        &   & V_{mp}
              \left(
              \frac{\left|
                    \omega(s(\xi,\nu_{mxt}),F_r(t)) s(\xi,\nu_{mxt}) -
                    \omega(s(\xi,\nu_{mxr}),F_r(t)) s(\xi,\nu_{mxr})
                    \right|}
                   {\omega(s(\xi,\nu_{mxt}),F_r(t)) s(\xi,\nu_{mxt})}
              \right) {\rm \, ,} \nonumber \\
        &   &   
\end{eqnarray}
where $V_{mp}$ is the expected order of magnitude for $V_{TOV}$. The relation between the coordinates $\nu_{mxt}$ and $\nu_{mxr}$ of a particular point has as follows 
\begin{equation}
\nu_{mxt}(\nu_{mxr},\nu_{\chi}) = \left \{ 
            \begin{array}{ll}
   + \nu_{\chi} \nu_{mxr} - \sqrt{1 - \nu_{\chi}^2} \sqrt{1 - \nu_{mxr}^2} &
                            {\rm if} \ \chi + \vartheta_{mxr} \leq 90 \arcdeg \\   
   - \nu_{\chi} \nu_{mxr} + \sqrt{1 - \nu_{\chi}^2} \sqrt{1 - \nu_{mxr}^2} &
                            {\rm if} \ \chi + \vartheta_{mxr} > 90 \arcdeg 
            \end{array}
            \right.
\end{equation}
Assuming for simplicity that the coordinate $\nu$ (without subscript) refers to the mixer configuration, we can write for the turn-over viscosity (G01,eq. [58]) 
\begin{equation}        
\mu_{TOV}(\xi,\nu,t) \simeq \frac{1}{3} \varrho(\xi) 
              A_{TOV}^v V_{TOV}(\xi,\nu,t) A_{TOV}^h h_P(\xi) {\rm \, .}   
\end{equation}
Then the energy dissipated over the time $t_{TOV}$, $D\!E_{TOV}$, is given by (G01, eq. [59])
\begin{eqnarray} \label{eq:tdottov}
D\!E_{TOV}&=& \int_0^{t_{TOV}}{ D_{TOV}(t) } \, dt = 4 \pi \alpha^3 \times 
              \nonumber \\ 
          & & \int_0^{t_{TOV}}{ \Omega_{3c}^2(t) \left( 
                    \int_0^1{
                    \int_0^{\xi_1}{ \mu_{TOV}(\xi,\nu,t) \, \xi^4 (1 - \nu^2) 
                \left( \frac{\partial \omega(s,F_r(t))}{\partial s} 
                \right)_{s=s(\xi,\nu)}^2 d\xi }
                              } d\nu \right)} dt {\rm \, ;} 
              \nonumber \\
          & &
\end{eqnarray}  
$D_{TOV}(t)$ is now a function of time, while in the foregoing interpretation it was taken approximately constant over $t_{TOV}$. The required equality $D\!E_{TOV} = D\!E_{DRD}$ is again the way for determining the fine-tuning parameter $A_{TOV}^v$.

On the other hand the energy, $D\!E_t$, dissipated over the time $t_{TOV}$ due to turbulent viscosity acting over a ``turbulent zone'' $[\xi_t,\xi_1]$ (lying, in turn, within the surface zone $[\xi_s,\xi_1]$) is given by
\begin{eqnarray} \label{eq:tdottvi}
D\!E_t&=& \int_0^{t_{TOV}}{ D_t(t) } \, dt = 4 \pi \alpha^3 \times 
              \nonumber \\ 
          & & \int_0^{t_{TOV}}{ \Omega_{3c}^2(t) \left( 
                    \int_0^1{
                    \int_{\xi_t}^{\xi_1}{ \mu_t(\xi,\nu,t) \, \xi^4 (1 - \nu^2) 
                \left( \frac{\partial \omega(s,F_r(t))}{\partial s} 
                \right)_{s=s(\xi,\nu)}^2 d\xi }
                                              } d\nu \right)} dt {\rm \, ,} 
              \nonumber \\
          & &
\end{eqnarray}
where $D_t(t)$ is the power dissipated due to turbulent viscosity. Consequently, the required equality $D\!E_{TOV} = D\!E_t$ can be used for accurately determining the base $\xi_t$ of the turbulent zone. In the following section, we shall use this method for several white dwarf and neutron star models. Actually, we shall determine the so-called ``representative turbulent zone'' (RTZ, with $\xi_t = \xi_{RTZ}$), which corresponds to an axisymmetric model with constant differential rotation, $F_r = F_{r[ini]}$, and angular velocity, $\Omega_{3c} = \Omega_{3c[ini]}$, suffering from power loss due to turbulent viscosity, $D_t$, equal to the average value (over the time $t_{TOV}$) of the power loss due to turn-over, $\left< D_{TOV} \right>_t$; that is, 
\begin{equation}
4 \pi \alpha^3 \Omega_{3c[ini]}^2 \, \int_0^1{\int_{\xi_{RTZ}}^{\xi_1} 
        { \mu_t(\xi,\nu,t) \, \xi^4 (1 - \nu^2) 
        \left( \frac{\partial \omega(s,F_{r[ini]})}{\partial s} 
        \right)_{s=s(\xi,\nu)}^2 d\xi }} d\nu = \left< D_{TOV} \right>_t 
        {\rm \, .}
\end{equation}
In order for this method to be reliable, however, the power dissipated due to electron and ion viscosity of degenerate matter (\cite{duri73}, \S \ 3, eqs. [4], [6], respectively), $D_{EIV}$ (calculated by eq. [\ref{eq:DTOV}] with $\mu_{TOV}$ substituted by the electron--ion viscosity $\mu_{e+i}$), should be small in comparison with the former, $D_t \gg D_{EIV}$; otherwise, the latter cannot be screened by the former and thus becomes significant in the framework of TOV scenario. 

\section{Some results}
In this paper, we present some preliminary results. In particular, Table 1 gives the most interesting quantities involved in the subject discussed above. Our computations concern 4 models of compact stars. Two of them simulate white dwarfs; the corresponding computations have been made by the so-called ``complex plane iterative technique'' (CIT; \cite[and references therein]{gpap00}). The other two models represent neutron stars; the corresponding computations should be considered as rough computations. An accurate numerical treatment of neutron stars is now in preparation.

As a summary, it seems that the representative turbulent zone is $\sim 1\%$ of the corresponding surface zone of these models; this fact shows the high efficiency of turbulent viscosity in dissipating the rotational kinetic energy due to differential rotation and, thus, leading the configuration from an unstable high energy level (aligned rotator) to a stable low energy level (perpendicular rotator) within a relatively small turn-over timescale.  

\acknowledgments{
{\bf Acknowledgments.} The research reported here was supported by the Research Committee of the University of Patras (C. Carath\'eodory's Research Project 1998/1932).}   

\clearpage

\begin{deluxetable}{lrrrr}
\scriptsize
\tablecolumns{5}
\tablewidth{0pt}
\tablecaption{Summary of calculations \label{tbl-1}}
\tablehead{
\colhead{} & \multicolumn{4}{c}{Model\tablenotemark{a} \ \& Mass\tablenotemark{b}} \\
\cline{2-5}                                                                        \\
\colhead{Parameter\tablenotemark{c}} 
& \colhead{WD\tablenotemark{d} \ , 0.60}  
& \colhead{WD, 0.89}
& \colhead{NS\tablenotemark{e} \ , 1.61}
& \colhead{NS, 2.39}
}
\startdata
Delta parameter (eqs. [8]--[10]), $\delta$ (dimensionless)
                                               & 1.00$(+00)$\tablenotemark{f} & 4.41$(+00)$ & 1.00$(+00)$ & 1.00$(+00)$
\nl
Initial turn-over angle, $\chi_{ini}$ (arcdegrees)
                                               & 1.41$(+00)$ & 1.41$(+00)$ & 1.32$(+00)$ & 1.41$(+00)$ 
\nl
Angular momentum, $L_{xx}$                     & 7.00$(+48)$ & 2.17$(+49)$ & 2.31$(+48)$ & 2.00$(+48)$
\nl
Average surface poloidal field, $B_s$          & 5.00$(+06)$ & 5.00$(+06)$ & 3.00$(+13)$ & 2.00$(+13)$
\nl
Magnetic flux, $f$                             & 4.01$(+24)$ & 2.28$(+24)$ & 4.44$(+25)$ & 1.90$(+25)$
\nl
TOV timescale, $t_{TOV}$ (yr)                  & 2.05$(+07)$ & 3.00$(+06)$ & 1.60$(+01)$ & 9.81$(+02)$
\nl
Present TOV time, $t_{now}$ (yr)               & 1.36$(+07)$ & 1.56$(+06)$ & 8.52$(+00)$ & 8.55$(+02)$
\nl
Present central period, $P_{now}$              & 1.42$(+02)$ & 3.31$(+01)$ & 2.14$(-03)$ & 2.55$(-03)$ 
\nl
Average spin-down time rate due to turn-over, $\left< \dot{P} \right>_t$
                                               & 8.90$(-14)$ & 1.68$(-13)$ & 1.67$(-12)$ & 3.07$(-14)$
\nl
Present turn-over angle, $\chi_{now}$ (arcdegrees)         
                                               & 7.11$(+01)$ & 7.21$(+01)$ & 6.30$(+01)$ & 8.31$(+01)$
\nl
Average power loss due to turn-over, $\left< D_{TOV} \right>_t$ 
                                               & 1.71$(+31)$ & 1.45$(+33)$ & 4.06$(+41)$ & 6.19$(+39)$
\nl
Power loss due to nonturbulent viscosity, $D_{EIV}$
                                               & 8.42$(+28)$ & 3.65$(+30)$ & 6.50$(+38)$ & 4.16$(+38)$
\nl 
Percent SZ\tablenotemark{g} \  thickness, $\frac{\xi_1-\xi_s}{\xi_1} \times 100$
                                               & 4.76$(+00)$ & 2.24$(+00)$ & 4.27$(-01)$ & 2.30$(-01)$
\nl
Percent RTZ\tablenotemark{h} \  thickness relative to SZ, $\frac{\xi_1-\xi_{RTZ}}{\xi_1-\xi_s} \times 100$
                                               & 5.32$(-01)$ & 2.15$(+00)$ & 5.47$(+00)$ & 2.52$(+00)$        
\nl
\cutinhead{Parameters of the axisymmetric differentially rotating starting model (aligned rotator)}
Central period, $P_{xx}$                       & 1.04$(+02)$ & 2.48$(+01)$ & 1.69$(-03)$ & 1.72$(-03)$
\nl
Rotational kinetic energy, $T_{xx}$            & 1.47$(+47)$ & 1.81$(+48)$ & 3.06$(+51)$ & 2.55$(+51)$
\nl
Moment of inertia along the $I_1$ axis, $I_{DR,11}(h,L_{xx})$
                                               & 1.79$(+50)$ & 1.39$(+50)$ & 9.15$(+44)$ & 8.45$(+44)$
\nl
Moment of inertia along the $I_3$ axis, $I_{DR,33}(h,L_{xx})$
                                               & 1.74$(+50)$ & 1.38$(+50)$ & 9.05$(+44)$ & 8.23$(+44)$
\nl
Average surface poloidal field, $B_s$          & 5.00$(+06)$ & 5.00$(+06)$ & 3.00$(+13)$ & 2.00$(+13)$
\nl
\cutinhead{Paprameters of the nonaxisymmetric rigidly rotating terminal model (perpendicular rotator)}
Period, $P_{RR}$                               & 1.62$(+02)$ & 4.07$(+01)$ & 2.54$(-03)$ & 2.67$(-03)$
\nl
Rotational kinetic energy, $T_{RR}$            & 1.36$(+47)$ & 1.67$(+48)$ & 2.86$(+51)$ & 2.35$(+51)$
\nl
Moment of inertia along the $I_1$ axis, $I_{RR,11}(h,L_{xx})$
                                               & 1.80$(+50)$ & 1.41$(+50)$ & 9.31$(+44)$ & 8.49$(+44)$
\nl
Moment of inertia along the $I_3$ axis, $I_{33}(h)$
                                               & 1.72$(+50)$ & 1.28$(+50)$ & 8.63$(+44)$ & 8.11$(+44)$
\nl     
Average surface poloidal field, $B_s$          & 5.05$(+06)$ & 5.27$(+06)$ & 3.10$(+13)$ & 2.02$(+13)$
\enddata
\tablenotetext{a}{Basic model: $F_r = 1.00$; $h = 0.08$}
\tablenotetext{b}{In solar masses, $M_{\sun}$} 
\tablenotetext{c}{In cgs units, unless stated otherwise}
\tablenotetext{d}{WD: white dwarf (Chandrasekhar's model), accurate computations} 
\tablenotetext{e}{NS: neutron star (Ideal Neutron Gas equation of state), rough computations} 
\tablenotetext{f}{Parenthesized numbers denote powers of 10}
\tablenotetext{g}{SZ: surface zone}
\tablenotetext{h}{RTZ: representative turbulent zone}
\end{deluxetable}

\end{document}